\documentclass{emulateapj}
\usepackage{apjfonts}
\usepackage{epsfig}
\usepackage{natbib}

\newcommand{\chisq}{\ensuremath{\chi^2}}

\newcommand{\flux}{erg s$^{-1}$ cm$^{-2}$}

\def\gtrsim{\mathrel{\hbox{\rlap{\hbox{\lower4pt\hbox{$\sim$}}}\hbox{\raise2pt\hbox{$>$}}}}}
\newcommand{\halpha}{H\ensuremath{\alpha}}
\newcommand{\hbeta}{H\ensuremath{\beta}}
\newcommand{\hn}{\halpha+[\ion{N}{2}]}

\newcommand{\kms}{km s\ensuremath{^{-1}}}

\newcommand{\lf}{\ensuremath{L_{5100}}}
\newcommand{\lum}{erg s$^{-1}$}

\newcommand{\mbh}{\ensuremath{M_\mathrm{BH}}}

\newcommand{\msigma}{\ensuremath{M_{\mathrm{BH}}-\sigmastar}}
\newcommand{\msun}{\ensuremath{M_{\odot}}}

\newcommand{\sigmastar}{\ensuremath{\sigma_{\star}}}

\newcommand{\sur}{$u-r$}

\slugcomment{To appear in {\it The Astrophysical Journal}}
\shorttitle{Intermediate-mass Black Holes}
\shortauthors{Greene \& Ho}

\begin{document}

\title{Active Galactic Nuclei with Candidate Intermediate-mass Black Holes}

\author{Jenny E. Greene}
\affil{Harvard-Smithsonian Center for Astrophysics, 60 Garden St., 
Cambridge, MA 02138}

\and

\author{Luis C. Ho}
\affil{The Observatories of the Carnegie Institution of Washington,
813 Santa Barbara St., Pasadena, CA 91101-1292}

\begin{abstract}

We present an initial sample of 19 intermediate-mass black hole
candidates in active galactic nuclei culled from the First Data
Release of the Sloan Digital Sky Survey.  Using the
linewidth-luminosity-mass scaling relation established for broad-line
active nuclei, we estimate black hole masses in the range of
$M_{\mathrm{BH}} \approx 8 \times 10^4 - 10^6 \, M_{\sun}$, a regime
in which only two objects are currently known. The absolute magnitudes
are faint for active galactic nuclei, ranging from $M_g \approx -15$
to $-18$ mag, while the bolometric luminosities are all close to the
Eddington limit.  The entire sample formally satisfies the
linewidth criterion for so-called narrow-line Seyfert 1 galaxies;
however, they
display a wider range of \ion{Fe}{2} and [\ion{O}{3}] $\lambda$5007
line strengths than is typically observed in this class of objects.
Although the available imaging data are of insufficient quality to
ascertain the detailed morphologies of the host galaxies, it is likely
that the majority of the hosts are relatively late-type systems.  The
host galaxies have estimated $g$-band luminosities $\sim 1$ mag
fainter than $M^*$ for the general galaxy population at $z \approx
0.1$.  Beyond simply extending the known mass range of central black
holes in galactic nuclei, these objects provide unique observational
constraints on the progenitors of supermassive black holes.  They are
also expected to contribute significantly to the integrated signal for
future gravitational wave experiments.

\end{abstract}


\section{Introduction}

Dynamical studies have established the existence of supermassive black
holes (BHs) with masses $M_{\rm BH} \approx 10^6-10^9$~$M_{\sun}$ in
the centers of most, if not all, local galaxies with bulges (Magorrian
et al.  1998; Richstone 2004).  A significant challenge to any model
of cosmological BH growth is the nature of seed BHs.  Observations of
intermediate-mass BHs ($M_{\rm BH} \approx 10^3-10^6$~$M_{\sun}$) in
the local Universe would provide the most direct empirical constraints
on such seeds.  The \msigma\ relation (Ferrarese \& Merritt 2000;
Gebhardt et al. 2000a; Tremaine et al.  2002) strongly indicates
feedback in the coevolution of galaxies and BHs, but it is still an
open question whether the relation is established at early or late
times.  If, as some models suggest (e.g., Di~Matteo et al. 2003), the
\msigma\ relation is established late in the evolution of a galaxy,
then we may expect to see deviations from this relation in a
population of intermediate-mass BHs that is not yet fully grown.
There are other motivations to characterize the lowest end of the
nuclear BH mass distribution.  Certainly their number and mass
distribution will impact the expected integrated background in
gravitational radiation (e.g., Hughes 2002).  Presumably they will
pass through an active phase of accretion and contribute at some level
to the observed integrated background radiation, particularly at X-ray
energies.  The range of accretion phenomena they display will depend
on the major accretion modes for BHs of such low mass.  The accretion
states and, by extension, radiative growth mechanisms of this
population will therefore have direct implications for the growth
history of primordial BHs.  Apart from their physical implications,
intermediate-mass BHs have practical value in that they
offer tremendous leverage in anchoring local BH-galaxy scaling 
correlations such as the \msigma\ relation.

Recently discovered intermediate-mass BH candidates in galactic nuclei
offer tantalizing hints as to the nature of this population.  NGC 4395
is a very late-type (Sdm) spiral with no bulge component, whose
central stellar velocity dispersion is \sigmastar$<30$ \kms\
(Filippenko \& Ho 2003).  Nevertheless, it has the emission properties
of a type 1 active galactic nucleus (AGN), with broad
permitted optical and UV emission lines (Filippenko \& Sargent 1989;
Filippenko, Ho, \& Sargent 1993), a compact radio core (Ho \& Ulvestad
2001) with a nonthermal brightness temperature (Wrobel, Fassnacht, \&
Ho 2001), and a pointlike, hard X-ray source (Ho et al. 2001) that is
highly variable (Moran et al. 1999, 2004; Shih, Iwasawa, \& Fabian
2003).  Mass estimates based on the \hbeta\ linewidth-luminosity-mass
scaling relation and X-ray variability suggest a BH mass of $10^4 -
10^5$~\msun\ (Filippenko \& Ho 2003).  Interestingly, these agree with
the limit of $< 10^5$~\msun\ derived from the \msigma\ relation
(Filippenko \& Ho 2003).  POX 52, on the other hand, has a dwarf
elliptical host.  Recently revisited by Barth et al. (2004), POX 52
was first observed by Kunth, Sargent, \& Bothun (1987), who noted the
Seyfert-like characteristics of its optical spectrum, including a
broad component to the \hbeta\ emission line.  The galaxy has a
central velocity dispersion of 36~\kms, which yields a mass estimate
of $M_{\rm BH} = 1.4 \times 10^5$~\msun, again consistent with the
value of $1.6 \times 10^5$ \msun\ derived from the \hbeta\
linewidth-luminosity-mass scaling relation.  The [\ion{O}{3}]
$\lambda$5007 linewidth, $\sigma~\approx~37$ \kms, is also fully
consistent with the stellar velocity dispersion (Barth et al. 2004).

Unfortunately, it is currently technically impossible to obtain direct
mass measurements for $M_{\rm BH} \lesssim 10^6$ \msun, because of our
inability to resolve the BH sphere of influence, for all but the
nearest galaxies in the Local Group (e.g., M33; Gebhardt et al. 2001).
The only direct mass determinations for intermediate-mass BHs come
from observations of massive star clusters (Gebhardt, Rich, \& Ho
2002; Gerssen et al. 2002), which, in any case, have been
controversial (Baumgardt et al. 2003a, b).  Barring evidence from
spatially resolved kinematics, the best indicator for nuclear BHs in
nearby galaxies comes from AGN activity as revealed through optical
emission-line intensity ratios (Baldwin, Phillips, \& Terlevich 1981;
Veilleux \& Osterbrock 1987; Ho, Filippenko, \& Sargent 1993),
kinematically distinct, broad emission lines (e.g., Ho et al. 1997c),
or the presence of compact, nonstellar radio and X-ray cores (e.g., Ho
et al. 2001; Nagar et al. 2002; Terashima \& Wilson 2003; see Ho 2004a
for a review).

Thus, the best hope to find intermediate-mass BHs is through an AGN
survey.  However, there are at least two practical challenges.  First,
it may be that objects like NGC 4395 are quite rare.  In the Palomar
survey of $\sim 500$ nearby galaxies, NGC 4395 is the only example of
its kind found (Ho, Filippenko, \& Sargent 1997a, b).  POX 52 was
discovered serendipitously in a fairly limited objective prism survey
for emission-line objects (Kunth et al.  1987).  These two are the
only known cases of intermediate-mass BHs in galactic nuclei.  Thus,
to accumulate a significant sample will require a much larger survey.
Second, even radiative signatures of accretion in these objects will
be challenging to uncover.  The host galaxies are expected to be
either very late-type systems with low mass, and therefore
intrinsically faint, or else regular, luminous disk-dominated galaxies
with a tiny or nonexistent bulge.  Even at a high accretion rate a
$10^5$~\msun\ BH will itself have a low luminosity.  As a result,
detection of the faint AGN will be limited by sensitivity and host
galaxy contamination.

A large-area, sensitive, and uniform optical spectroscopic galaxy
survey, such as the Sloan Digital Sky Survey (SDSS; York et~al. 2000), 
offers the best
opportunity for finding a significant number of new intermediate-mass
BH candidates.  We present the first sample of candidates culled from
the SDSS.  We discuss the data in \S 2, the sample definition and
analysis in \S 3, and the main results in \S 4.  In this paper we
assume the following cosmological parameters to derive distances: $H_0
= 100h = 72$~\kms~Mpc$^{-1}$, $\Omega_{\rm m} = 0.3$, and
$\Omega_{\Lambda} = 0.7$.

\section{The SDSS Data}

Since we have relied heavily on the SDSS software pipeline, we will
briefly discuss relevant characteristics of the survey.  This is by no
means a complete discussion.  Details of SDSS photometry and
spectroscopy can be found in the Early Data Release paper (Stoughton
et al.  2002; hereafter EDR) and the Data Release One paper (Abazajian
et al. 2003; hereafter DR1).

The SDSS uses a dedicated 2.5~m telescope to image one-quarter of the
sky and obtain follow-up spectroscopy on automatically selected star,
galaxy, and quasar candidates.  The DR1 includes a photometric sample of 53 
million objects over 2099 deg$^2$ and photometrically defined quasar, galaxy,
and stellar spectra over 1360 deg$^2$ (Abazajian et al. 2003).

Calibration and sample selection for both surveys is conducted 
by automated pipelines.  The photometric pipeline, \emph{Photo}, 
performs bias subtraction, flat-fielding, cosmic ray rejection, 
and masking of bad pixels, followed by astrometry,
point-spread function modeling, object identification, and image deblending.
It measures four different (asinh) magnitudes (Lupton, Gunn, \& Szalay 1999).
The Petrosian magnitude, measured within a radius defined by the shape
of the azimuthally averaged light profile of the galaxy, is best for
photometry of extended objects such as galaxies (Stoughton et al.  2002).
Typical photometric calibration errors are 0.02 mag in $g$ (Smith et al.  
2002), while the zeropoints between the SDSS and AB magnitude system could 
differ by as much as $10\%$ (Stoughton et al.  2002).

Galaxy targets are automatically selected based on morphology.  Specifically,
resolved sources with $r \lesssim 17.77$ mag are targeted as galaxies (note 
that this cut-off varied between the EDR and the DR1).  Candidates with 
magnitudes brighter than 15 in $g$ or $r$ or 14.5 in $i$
are rejected in case of contamination of other fibers.  A surface 
brightness cut is also applied, and targets may be excluded due to
proximity because the fiber holders have a finite size
and are kept $55\arcsec$ apart from each other to avoid fiber collisions.
Quasar candidates are selected on the
basis of a color selection technique detailed in Richards et al. (2002).

SDSS spectra are acquired with a pair of double, fiber-fed spectrographs.  A 
plate of $3~\mathrm{deg}^2$ with 640 optical fibers is drilled for each 
field.  Each fiber subtends a diameter of $3\arcsec$, corresponding to 
$\sim 6.5$~kpc at $z=0.1$.  The spectra have an instrumental resolution of 
$\lambda/\Delta \lambda \approx 1800$ ($\sigma_{\rm ins} \approx 71$ \kms).  
Integration times are determined for a minimum signal-to-noise ratio of 4 at 
$g=20.2$ mag.  The spectroscopic pipeline performs basic calibration as above, 
as well as spectral extraction, sky subtraction, removal of atmospheric 
absorption bands, and wavelength and spectrophotometric calibration.
Redshifts are determined using \chisq\ fits to stellar, emission-line galaxy,
and quasar templates, as well as to the emission lines themselves; the 
typical accuracy is $\sim 30$~\kms.


\begin{deluxetable}{rclccc}
\tablecolumns{6} 
\tabletypesize{\scriptsize}
\tablewidth{0pc}
\tablecaption{The SDSS Sample \label{tablebasic}}
\tablehead{ 
\colhead{ID} & \colhead{SDSS Name} & \colhead{$z$} & \colhead{$g$} & 
\colhead{$g-r$} & \colhead{$A_{g}$} \\
\colhead{(1)} & \colhead{(2)} & \colhead{(3)} &
\colhead{(4)} & \colhead{(5)} & \colhead{(6)}
}
 
\startdata

1 & SDSS J010712.03+140844.9 & 0.0768 & 18.49 & 0.62 & 0.26\\
2 & SDSS J024912.86$-$081525.6 & 0.0295 & 16.55 & 0.66 & 0.11\\
3 & SDSS J032515.59+003408.4 & 0.102 & 19.51 & 0.70 & 0.49\\
4 & SDSS J082912.67+500652.3 & 0.0434 & 17.68 & 0.41 & 0.16\\
5 & SDSS J094310.12+604559.1 & 0.0742 & 18.32 & 0.57 & 0.09\\
6 & SDSS J101108.40+002908.7 & 0.100 & 19.39 & 0.56 & 0.13\\
7 & SDSS J101627.32$-$000714.5 & 0.0943 & 19.31 & 0.66 & 0.13\\
8 & SDSS J114008.71+030711.4 & 0.0811 & 17.32 & 0.51 & 0.07\\
9 & SDSS J115138.24+004946.4 & 0.194 & 19.02 & 1.20 & 0.10\\
10 & SDSS J124035.81$-$002919.4 & 0.0809 & 18.00 & 0.47 & 0.09\\
11 & SDSS J125055.28$-$015556.6 & 0.0814 & 18.23 & 0.37 & 0.08\\
12 & SDSS J135724.52+652505.8 & 0.106 & 18.45 & 0.36 & 0.07\\
13 & SDSS J141234.67$-$003500.0 & 0.126 & 18.19 & 0.59 & 0.17\\
14 & SDSS J143450.62+033842.5 & 0.0281 & 15.60 & 0.62 & 0.14\\
15 & SDSS J144507.30+593649.9 & 0.127 & 19.98 & 0.41 & 0.03\\
16 & SDSS J170246.09+602818.9 & 0.0690 & 18.22 & 0.59 & 0.07\\
17 & SDSS J172759.15+542147.0 & 0.0994 & 19.27 & 0.57 & 0.14\\
18 & SDSS J232159.06+000738.8 & 0.183 & 19.39 & 0.71 & 0.16\\
19 & SDSS J233837.10$-$002810.3 & 0.0355 & 16.87 & 0.60 & 0.11\\

\enddata
\tablecomments{
Col. (1): Identification number assigned in this paper. 
Col. (2): Official SDSS name. 
Col. (3): Redshift measured by the SDSS pipeline.
Col. (4): Petrosian $g$ magnitude.  
Col. (5): Petrosian $g-r$ color.  
Col. (6): Galactic extinction in the $g$ band.  
}
\end{deluxetable}



\begin{figure*}
\begin{center}
\epsfig{file=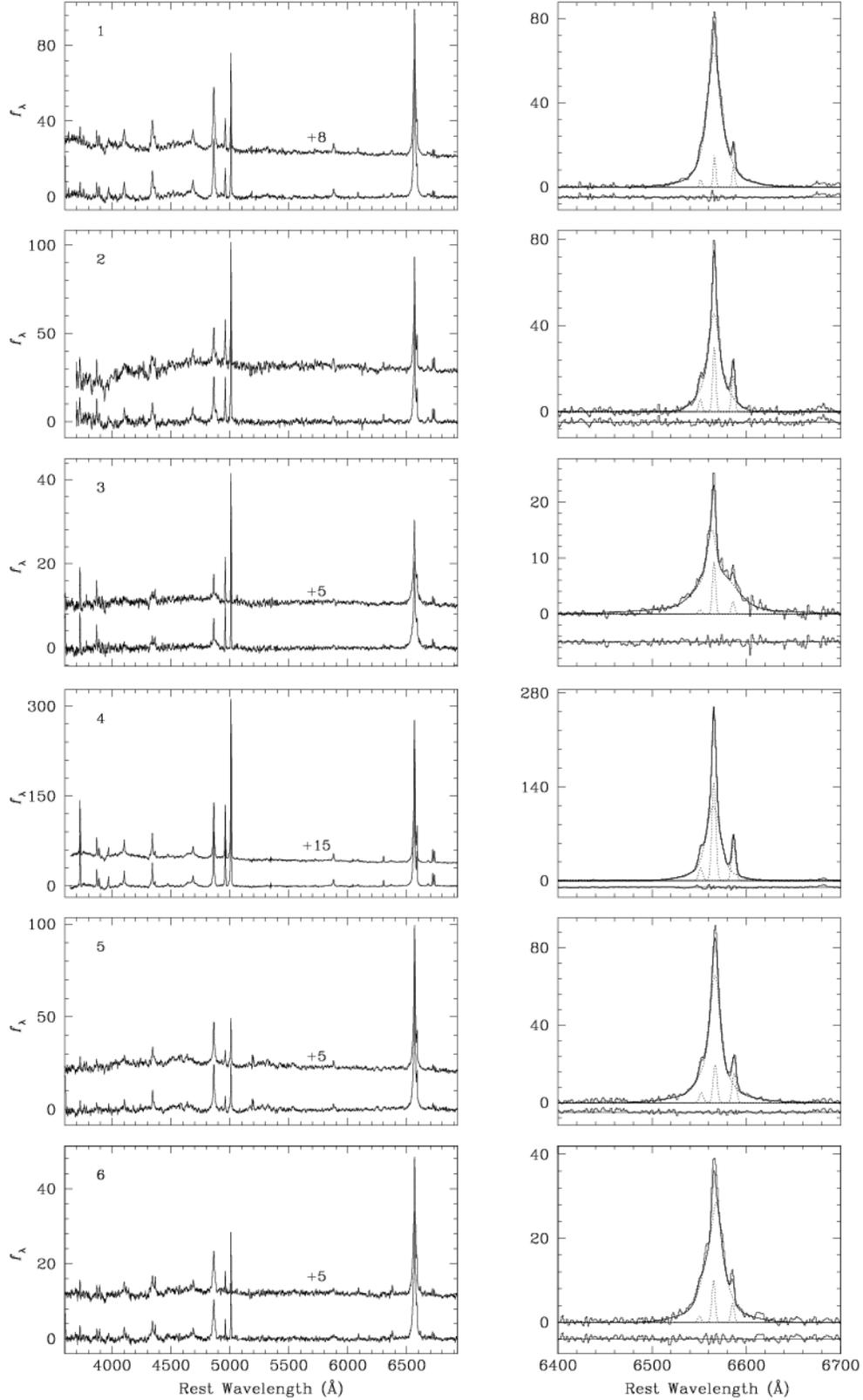,width=0.95\textwidth,totalheight=8.5in,keepaspectratio=true,angle=180}

\vskip -0.1mm
\figcaption[fig1.ps]{ 
{\it Left:} Spectrum with (bottom) and without
(top) starlight and continuum subtraction.  For clarity, the top
spectrum in some cases is offset by an additive constant, which is
indicated.  The top left corner of each panel gives the object
identification number, which can be referenced in Table
1.  {\it Right:} Enlargement of the region around
\halpha+[N~{\tiny II}] $\lambda\lambda$6548, 6583.  In the upper plot,
the data are shown in histogram, the individual models for the narrow
and broad lines are shown in dotted lines, and the sum of all the
fitted components is shown as a solid line.  The lower plot gives the
residuals of the fit.  The ordinate of the plots are in units of
$10^{-17}$ erg s$^{-1}$ cm$^{-2}$ \AA$^{-1}$.
\label{fig1}}
\vskip -5mm
\end{center}
\end{figure*}



\begin{figure*}
\begin{center}
\epsfig{file=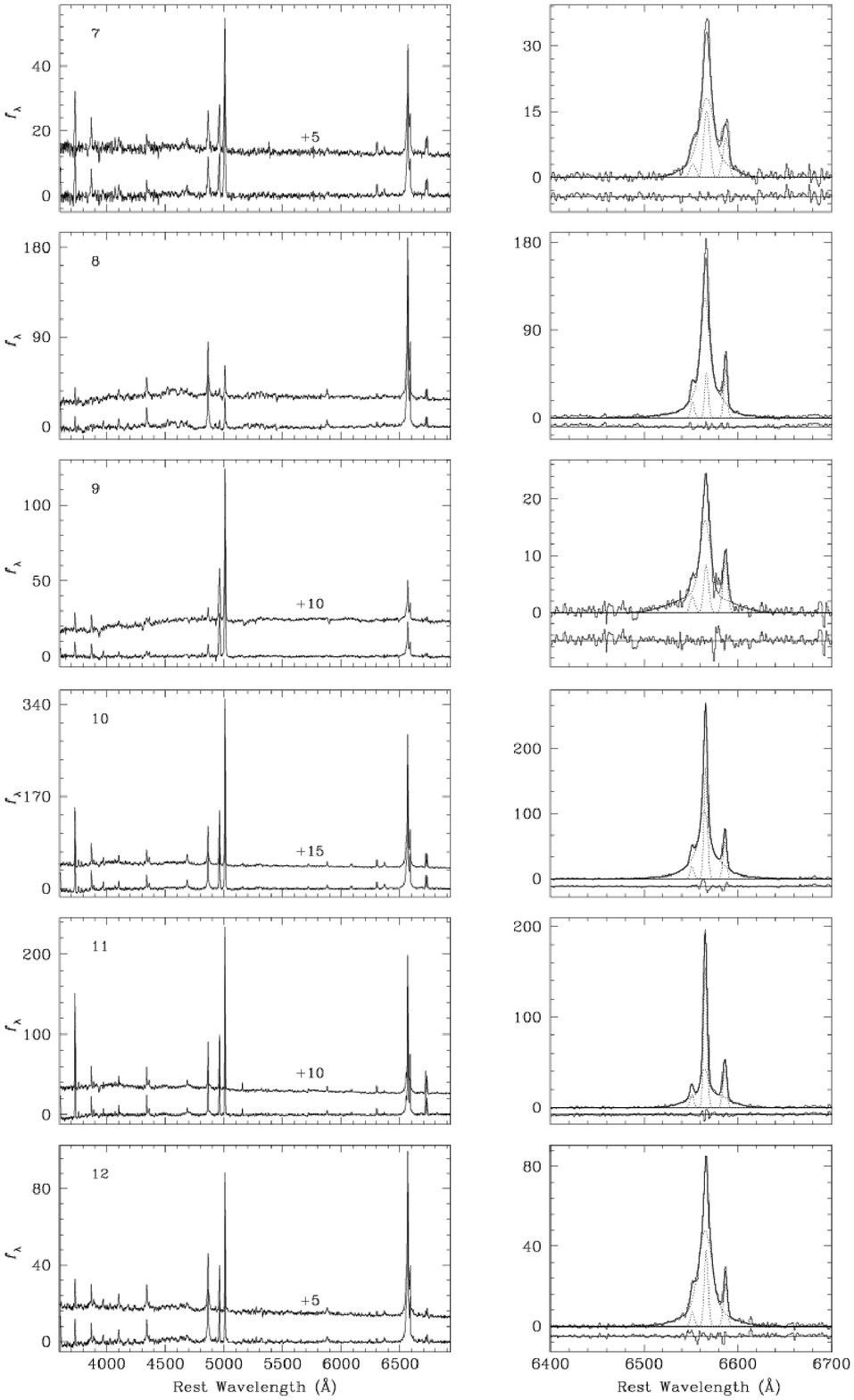,width=0.95\textwidth,totalheight=8.5in,keepaspectratio=true,angle=180}

\vskip -0.1mm
\figcaption[f2.eps]{ 
Same as Figure 1.
\label{fig2}}
\vskip -5mm
\end{center}
\end{figure*}



\begin{figure*}
\begin{center}
\epsfig{file=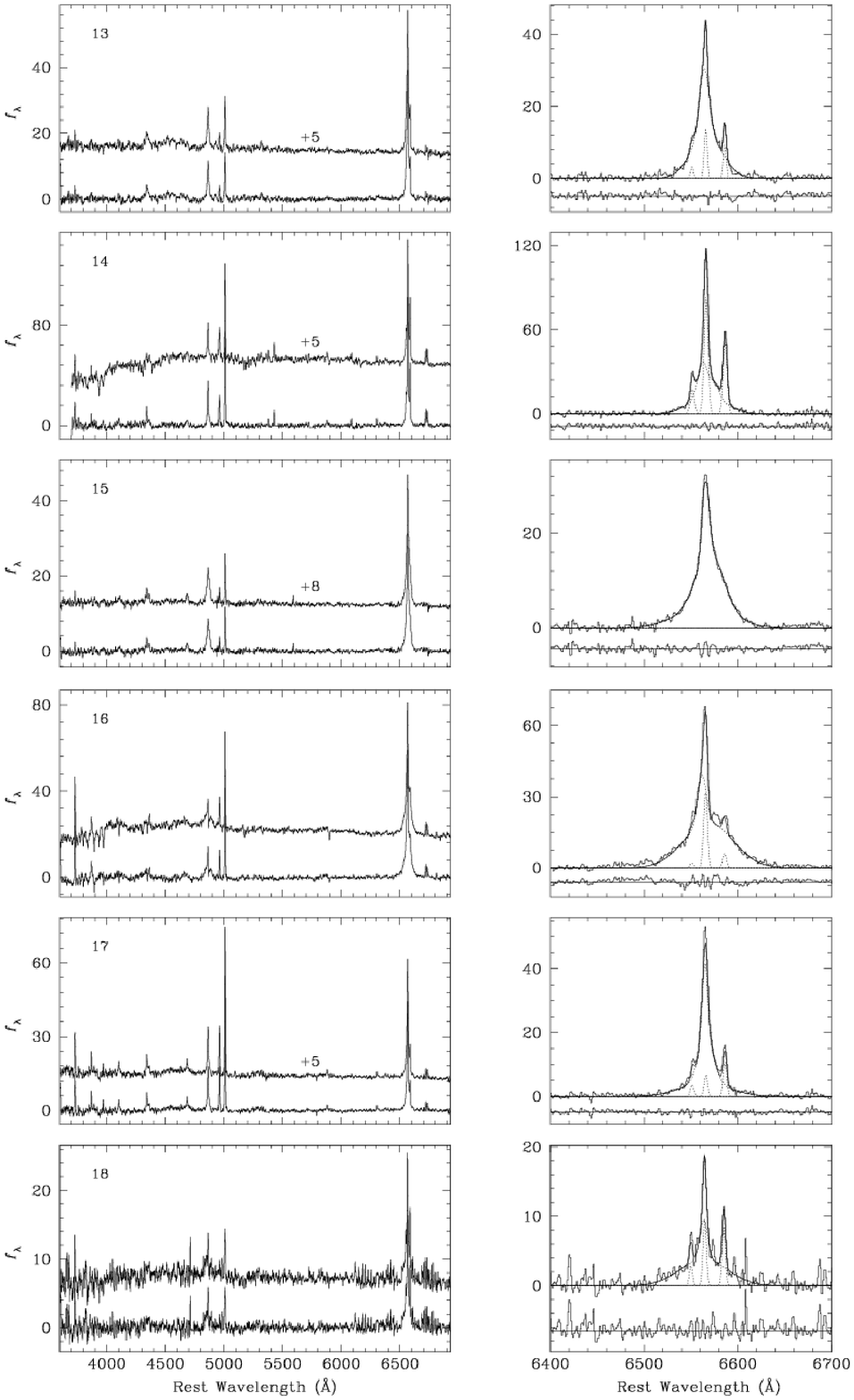,width=0.95\textwidth,totalheight=8.5in,keepaspectratio=true,angle=180}
\vskip -0.1mm
\figcaption[f3.eps]{ 
Same as Figure 1.
\label{fig3}}
\end{center}
\end{figure*}

\clearpage


\begin{figure*}
\plotone{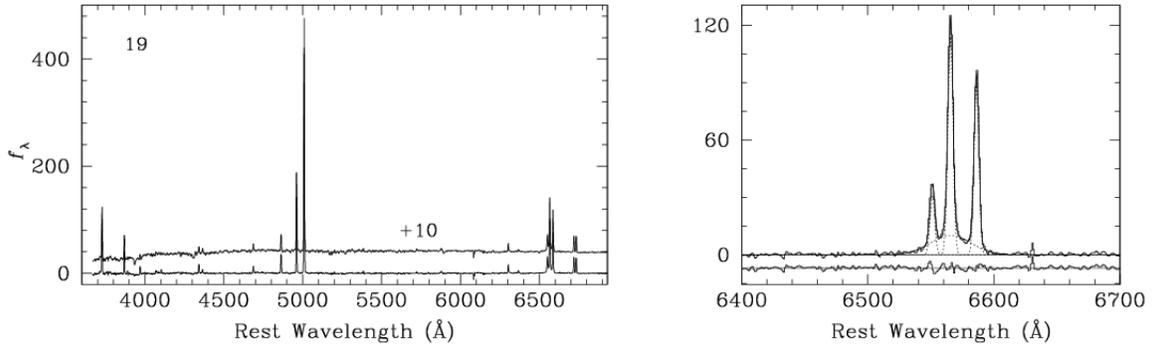}
\caption{ 
Same as Figure 1.
\label{fig4}}
\vskip -10mm
\end{figure*}


\section{The Sample and Analysis}

Our primary sample was culled from a parent sample of broad-line AGNs
at $z \leq 0.35$ in the DR1.  These are unambiguous AGN candidates
with virial BH mass estimates $\leq 10^6$~\msun, which hereafter we
consider to be the upper bound for intermediate-mass BHs.  We
determine ``virial'' BH masses using a variant of the
linewidth-luminosity-mass scaling relation of Kaspi et al. (2000),
whose robustness has been tested with respect to the \msigma\ relation
of inactive galaxies (Gebhardt et al. 2000b; Ferrarese et al. 2001;
see Barth 2004 for a review).  In order to apply this method, we first
identify all the AGNs from DR1 that have a broad component to at least
the H$\alpha$ emission line.  In order to detect objects with weak
lines and to reliably measure emission-line and continuum parameters,
we must properly remove the contaminating starlight from the host
galaxy, model the profiles of the emission lines, and quantify the
strength of the AGN continuum luminosity.  We discuss these various
steps below.

\subsection{Starlight and Continuum Subtraction}

Low-luminosity AGNs are challenging to detect because the signature emission
lines are often contaminated by starlight.  The SDSS fibers always include
significant host galaxy emission, making careful removal of the stellar
continuum essential to reliably measure the emission-line spectrum, and
especially to detect any weak, broad features.  Because we expect H$\alpha$
to be the strongest permitted line in the optical, and hence best suited
for detecting a broad-line component if present, we start with the
$\sim 153,000$ galaxies and quasars in the DR1 with $z \leq 0.35$.

The next step is to remove the starlight, along with any additional
featureless nonstellar continuum emission, in order to produce a pure
emission-line spectrum.  A variety of template-fitting techniques
exist to accomplish this (see, e.g., Ho 2004a for a review).  For this
work we use a principal component analysis (PCA) algorithm adopted
from Hao \& Strauss (2004; see also Connolly \& Szalay 1999).
Assuming a linear relation between all spectral variations, PCA (or
the Karhunen-Lo\'{e}ve transform) transforms a large library of input
spectra to an orthogonal basis in which each eigenspectrum is along a
direction of maximal variance.  The input spectra are derived from
absorption-line spectra from the SDSS, and an A-star spectrum may be
included to account for a post-starburst stellar population.  Each
program spectrum is then modeled as a linear combination of the
eigenspectra.  The PCA technique has a number of benefits when a
large, homogeneous database like SDSS is available.  The method
typically requires no more than eight eigenspectra to fully
reconstruct a given galaxy spectrum, and for absorption-line galaxies
the resulting solution is unique.  Furthermore, the eigenvectors
naturally include variance in velocity dispersion as long as the
template galaxies span the same range as the target sample.  In
practice, some galaxies contain substantial nonstellar continuum
emission, which we model as a power law fixed to a slope of $\beta =
1.0$, where $f_{\lambda} \propto \lambda^{-\beta}$.  
As discussed in J. E. Greene \& L. C. Ho (in preparation),
the AGN-dominated objects in our sample typically have a
power-law continuum spectrum with this slope.

Examples of spectra before and after starlight and continuum 
subtraction are given in the left panels of Figures 1--4.

\subsection{Initial Selection of Broad \halpha\ Candidates}

Once we obtained continuum-subtracted spectra, we used, as a first
pass, a simple algorithm to cull objects with excess emission in the
region 6300--6700 \AA\ that is potentially broad \halpha\
emission.  We began by removing any narrow emission lines from this
spectral region by adaptively smoothing it, iteratively removing
outlier points.  We then calculated the mean, root mean square (rms)
normalized to the dispersion in a line-free spectral region (5600-6000
\AA), and integrated flux of the region.  Since the spectra are
continuum-subtracted, any object with a nonzero mean, rms, or
integrated flux is a candidate for having broad \halpha\ emission.  By
trial and error we chose thresholds in the mean, integrated flux, and
normalized rms that identified all broad-line objects down to a very
low contrast.  We conservatively chose a low threshold, which
increased our contamination rate but enhanced our completeness.  The
resulting sample from this procedure comprised 4163 candidates.  As an
initial cut, this technique has the benefit of being computationally
inexpensive (important for an initial sample as large as the DR1).  It
has two major disadvantages: potential incompleteness and high
contamination.  Contaminants included misclassified late-type stars,
which tend to have a broad maximum near 6500 \AA\ that can be mistaken
for broad \halpha\ emission (Ho et al. 1997c), and objects of unknown
origin with large calibration errors or whose spectra have been
corrupted in the \halpha\ region.  We experimented with using pipeline
flags to automatically remove these, but in the end flagged them by
hand.  A more significant contaminant comes from incomplete removal of
narrow emission lines from starburst galaxies with extremely strong
line emission, or emission-line galaxies with multi-component,
relatively broad narrow lines.  Detailed profile fitting, 
as described below, is required to isolate galaxies 
with a truly kinematically distinct broad-line component.

\subsection{Emission-line Profile Fitting}

The broad \halpha\ feature, a reliable signature of AGN activity (see
Ho et al. 1997c), is coincident with a combination of narrow \halpha\
emission and the flanking [\ion{N}{2}] $\lambda \lambda$6548, 6583
doublet.  While in many cases the broad-line emission in our
candidates is evident by eye, the \hn\ region is sufficiently
complicated that some care is needed to determine the presence or
absence of a broad \halpha\ component.  In addition to imperfect
removal of starlight, the shape of the narrow lines themselves may
confuse the detection of broad \halpha.  There are often broad,
non-Gaussian wings, multiple peaks, and asymmetries in the narrow
lines that can mimic or hide a true broad component to \halpha.  We
therefore use a profile-fitting procedure, similar to that of Ho
et al. (1997c), to model the narrow components based on regions of the
spectrum less prone to blending and confusion.  We can then determine
objectively whether a broad component is present.

While the profiles of the narrow emission lines in some AGNs are known
to vary as a function of critical density and ionization parameter
(e.g., Pelat, Alloin, \& Fosbury 1981; Filippenko \& Halpern 1984;
Filippenko 1985; De~Robertis \& Osterbrock 1986), empirically the line
profile of [\ion{S}{2}] $\lambda \lambda$6716, 6731 is generally well
matched to those of [\ion{N}{2}] and narrow \halpha\ (Filippenko \&
Sargent 1988; Ho et al. 1997c).  Therefore, we model the [\ion{S}{2}]
doublet with a superposition of Gaussian components and then scale
this model to fit the \hn\ narrow lines.  We use up to four Gaussian
components when it is statistically justified.  We do not ascribe any
particular physical significance to the individual components, but
they allow us to describe the line shape as fully as the data quality
allows.  The two [\ion{S}{2}] lines are assumed to have the same
profile and are fixed in separation by their laboratory wavelengths; the
relative strengths of the two lines change with density and so are
allowed to vary in the fit. In a few cases where [\ion{S}{2}] is
undetectable or too weak to yield a reliable fit, we construct a model
using [\ion{O}{3}] $\lambda 5007$.  [\ion{O}{3}] is both strong and
nearly ubiquitous, and may seem like a better choice for template
fitting.  Unlike [\ion{S}{2}], the [\ion{O}{3}] profile does not
typically match that of [\ion{N}{2}] in detail.  
[\ion{O}{3}] often has a broad,
blue shoulder to its profile, 
suggestive of an outflow origin (De~Robertis \& 
Osterbrock 1984, 1986; Whittle 1985).  Nonetheless,
[\ion{O}{3}] still seems to be an acceptable substitute for
[\ion{S}{2}] when the latter is unavailable.

Once a satisfactory narrow-line model is achieved it serves as a
template for fitting the \hn\ complex.  The centroids of each
component are fixed relative to each other at their laboratory
separations, and the relative strength of [\ion{N}{2}] $\lambda 6583$
and [\ion{N}{2}] $\lambda 6548$ is fixed at the theoretical value of
2.96.  An additional broad component for \halpha\ is added if
statistically warranted; we use as many Gaussians as needed, but
generally no more than 2--3, to fully describe the complex shapes of
some of the broad components.  Our fits are shown as dotted lines in
the right panels of Figures 1--4.  From the final model for the broad
\halpha\ line, we measure the full width at half maximum (FWHM) of the
profile and its integrated flux.  This step eliminated a large number
of broad \halpha\ candidates, leaving a sample of $\sim$3200
genuinely broad-line objects.

\begin{deluxetable*}{rccccccccccccc}
\tablecolumns{14} 
\tablewidth{0pc}
\tabletypesize{\scriptsize}
\tablecaption{Emission-line Measurements \label{tableline}}
\tablehead{ 
\colhead{ID} & 
\colhead{[O {\tiny II}]} & 
\colhead{Fe {\tiny II}} & 
\colhead{(\hbeta)$_{\rm n}$} & 
\colhead{(\hbeta)$_{\rm b}$} & 
\colhead{[O {\tiny III}]} & 
\colhead{[O {\tiny I}]} & 
\colhead{(\halpha)$_{\rm n}$} &
\colhead{(\halpha)$_{\rm b}$} & 
\colhead{[N {\tiny II}]} &
\colhead{[S {\tiny II}]} &
\colhead{[S {\tiny II}]} & 
\colhead{FWHM$_{\mathrm{H\alpha}}$} & 
\colhead{FWHM$_{[\mathrm{O {\tiny III}}]}$} \\
\colhead{} & 
\colhead{$\lambda3727$} & 
\colhead{$\lambda4570$} & 
\colhead{} & 
\colhead{} & 
\colhead{$\lambda5007$} &
\colhead{$\lambda6300$} & 
\colhead{} & 
\colhead{} & 
\colhead{$\lambda6583$} &
\colhead{$\lambda6716$} & 
\colhead{$\lambda6731$} &
\colhead{(\kms)} & 
\colhead{(\kms)} \\
\colhead{(1)} & 
\colhead{(2)} & 
\colhead{(3)} & 
\colhead{(4)} & 
\colhead{(5)} & 
\colhead{(6)} & 
\colhead{(7)} &
\colhead{(8)} & 
\colhead{(9)} & 
\colhead{(10)} & 
\colhead{(11)} & 
\colhead{(12)} &
\colhead{(13)} &
\colhead{(14)} 
}
 
\startdata

1 &\quad0.154 & \quad1.59 & \quad0.046 &1.48 & \quad362. &\quad0.021 &0.143 &\quad4.73 &
\quad0.098 &\quad0.043 &\quad0.043 & \quad830 &\quad260\\
 
2 &\quad0.273 & $<$1.04 & \quad0.079 &0.88 & \quad519. &\quad0.053 &0.245 &\quad1.95 &
\quad0.141 &\quad0.077 &\quad0.074 & \quad732 &\quad262\\
 
3 &\quad0.225 & \quad1.07 & \quad0.070 &0.73 & \quad192. &$<$0.002 &0.217 &\quad3.25 &
\quad0.051 &\quad0.064 &\quad0.038 & \quad916 &\quad199\\
 
4 &\quad0.307 & \quad0.26 & \quad0.139 &0.42 & \quad1880. &\quad0.030 &0.431 &\quad1.48 &
\quad0.165 &\quad0.073 &\quad0.063 & \quad870 &\quad238\\
 
5 &\quad0.081 & \quad4.29 & \quad0.167 &1.79 & \quad210. &$<$0.005 &0.517 &\quad7.98 &
\quad0.400 &\quad0.097 &\quad0.055 & \quad698 &\quad291\\
 
6 &\quad0.171 & \quad2.83 & \quad0.074 &1.93 & \quad104. &$<$0.008 &0.461 &\quad8.31 &
\quad0.206 &\quad0.099 &\quad0.067 & \quad919 &\quad261\\
 
7 &\quad0.309 & \quad0.55 & \quad0.093 &0.25 & \quad453. &\quad0.081 &0.288 &\quad1.16 &
\quad0.170 &\quad0.090 &\quad0.095 & \quad940 &\quad449\\
 
8 &\quad0.096 & \quad6.98 & \quad0.311 &1.89 & \quad241. &\quad0.182 &0.966 &\quad11.0 &
\quad1.00 &\quad0.242 &\quad0.241 & \quad591 &\quad297\\
 
9 &\quad0.085 & $<$0.55 & \quad0.019 &0.07 & \quad1050. &$<$0.001 &0.060 &\quad0.46 &
\quad0.062 &\quad0.014 &\quad0.017 & \quad703 &\quad504\\
 
10 &\quad0.240 & \quad0.41 & \quad0.127 &0.25 & \quad2150. &\quad0.048 &0.396 &\quad1.35 &
\quad0.138 &\quad0.071 &\quad0.069 & \quad722 &\quad255\\
 
11 &\quad0.578 & \quad0.44 & \quad0.219 &0.20 & \quad1140. &\quad0.043 &0.680 &\quad1.01 &
\quad0.203 &\quad0.155 &\quad0.118 & \quad642 &\quad212\\
 
12 &\quad0.133 & \quad0.92 & \quad0.154 &0.75 & \quad463. &\quad0.020 &0.478 &\quad2.87 &
\quad0.263 &\quad0.040 &\quad0.054 & \quad872 &\quad231\\
 
13 &\quad0.191 & \quad3.55 & \quad0.157 &1.58 & \quad126. &$<$0.007 &0.487 &\quad6.65 &
\quad0.335 &\quad0.114 &\quad0.074 & \quad785 &\quad275\\
 
14 &\quad0.219 & $<$1.07 & \quad0.273 &0.52 & \quad541. &\quad0.053 &0.848 &\quad1.96 &
\quad0.523 &\quad0.159 &\quad0.140 & \quad1089 &\quad276\\
 
15 &\quad0.079 & \quad2.12 & \quad0.012 &2.93 & \quad83.4 &$<$0.004 &$<$0.038 &\quad11.8 &
$<$0.038 &$<$0.061 &$<$0.061 & \quad750 &\quad219\\
 
16 &\quad0.565 & $<$1.53 & \quad0.211 &1.18 & \quad257. &\quad0.028 &0.654 &\quad6.17 &
\quad0.124 &\quad0.148 &\quad0.115 & \quad919 &\quad234\\
 
17 &\quad0.244 & \quad0.97 & \quad0.028 &0.65 & \quad386. &\quad0.033 &0.086 &\quad2.08 &
\quad0.146 &\quad0.059 &\quad0.051 & \quad464 &\quad233\\
 
18 &\quad0.707 & $<$7.30 & \quad0.325 &0.75 & \quad49.5 &$<$0.038 &1.01 &\quad7.24 &
\quad0.899 &$<$0.434 &$<$0.434 & \quad707 &\quad342\\
 
19 &\quad0.234 & $<$0.13 & \quad0.075 & \nodata & \quad2550. &\quad0.045 &0.232 &\quad0.16 &
\quad0.175 &\quad0.072 &\quad0.069 & \quad1730 &\quad216\\

\enddata 
\tablecomments{
Col. (1): Identification number assigned in this paper. 
Col. (2)--(12): All fluxes are relative to that of 
[O {\tiny III}] $\lambda 5007$, which is listed in units of
$10^{-17}\mathrm{erg~s^{-1}~cm^{-2}}$.  Note that these are observed values; 
no extinction correction has been applied.  The subscripts ``n'' and ``b'' 
in Col. (4)--(5) and (8)--(9) refer to the narrow and broad components of
the line, respectively.  The narrow component of
\hbeta\ is sufficiently weak that it was impossible to obtain a robust
measurement of its strength, so we fixed it
to that of the narrow \halpha\ component,
assuming an intrinsic ratio of \halpha/\hbeta\ = 3.1 for AGNs
(Halpern \& Steiner 1983; Gaskell \& Ferland 1984).
Col. (13)--(14): These are the observed linewidths; they have not been 
corrected for instrumental resolution. The H$\alpha$ FWHM was derived 
from the model fit to the broad H$\alpha$ line, 
as described in \S 3.3.
}

\end{deluxetable*}


\subsection{Nonstellar Continuum Luminosity}

As discussed below, in order to estimate BH masses we need to know
the luminosity of the featureless nonstellar (AGN) continuum,
$L_{5100} \equiv \lambda L_{\lambda}$ at $\lambda$ = 5100 \AA.  We
obtain \lf\ in two ways.  For objects whose AGN component clearly
dominates (e.g., object \#1, 5, and 6 in Fig.  1), we directly fit a
single power law to the spectrum using spectral regions known to be
relatively unaffected by strong emission lines, especially from
\ion{Fe}{2}, which is pervasive in AGN spectra and can often be
mistaken for continuum emission.  The ``continuum'' windows we chose,
guided by Francis et al.  (1991), are: 3700--3715, 3740--3800,
4041--4043, 4150--4250, 5550--5850, 6000--6290, and 6400--6450 \AA.
From the power-law fit we derive a slope and the continuum flux at
$5100$ \AA.

For spectra with noticeable galaxy contribution, we must rely on the
PCA decomposition to model the power-law contribution. From extensive 
experimentation with the PCA fits, we find that we can reliably 
recover the power-law component in objects where the AGN contribution 
at 5100 \AA\ is $\gtrsim 20\%$.  In this study we only consider objects 
whose \lf\ exceeds this threshold.

Once we have a measurement of the AGN power-law continuum, we can
translate it into an equivalent $g$-band absolute magnitude,
$M_g$(AGN), by convolving the power law with the response function of
the SDSS $g$ filter.  In turn, we can estimate the $g$-band absolute
magnitude of the host galaxy, $M_g$(host), by subtracting the AGN
contribution to the total observed $g$-band luminosity.

\subsection{Black Hole Masses}

We estimate BH ``virial'' masses using a variant of the 
linewidth-luminosity-mass scaling relation discussed by Kaspi et al. (2000).
If the broad-line region (BLR) is in virial equilibrium, then the mass
of the central BH is \mbh$\approx \upsilon^2 R_{\rm BLR}/G$, where
$\upsilon = (\sqrt{3}/2) \times \upsilon_{\mathrm{FWHM}}$ and
$\upsilon_{\mathrm{FWHM}}$ is the FWHM of the broad component of
\hbeta, $R_{\rm BLR}$ is the radius of the BLR, and $G$ is the
gravitational constant.  For a sample of 34 low-redshift type 1 AGNs
studied with reverberation mapping Kaspi et al. (2000) found an
empirical relation between the continuum luminosity of the AGN at
$5100$ \AA\ and the size of the BLR:

\begin{equation}
R_{\rm BLR} = (32.9^{+2.0}_{-1.9}) \left[ \frac{{L_{5100}}}{10^{44}\, 
\mathrm{erg~s^{-1}}}\right]^{0.700 \pm 0.033} \mathrm{lt-days}.
\end{equation}

\noindent
With this scaling relation, the virial mass is simply

\begin{equation}
M_{\rm BH} = 4.82\times 10^6 
\left( \frac{{L_{5100}}}{10^{44}\, \mathrm{erg~s^{-1}}}\right)^{0.7}
\left( \frac{\upsilon_{\rm FWHM}}{1000\, \mathrm{km~s^{-1}}}\right)^{2} 
\, M_\odot.
\end{equation}

While it is conventional to use \hbeta\ to determine $\upsilon_{\rm FWHM}$, 
we have decided to use \halpha\ as a surrogate instead.  H$\alpha$  is at least
a factor of 3 stronger than \hbeta, and therefore much easier to measure 
for low-luminosity sources or in objects where host galaxy contamination is 
severe.  Another advantage of \halpha\ is that, despite blending with 
[\ion{N}{2}] emission, it is still much easier to measure because it is not 
strongly affected by nearby \ion{Fe}{2} emission, as is the case for \hbeta\ 
(see, e.g., Francis et al. 1991).

The currently popular formalism to derive virial BH masses for broad-line 
AGNs is calibrated against the sample of AGNs with reverberation mapping data 
(Kaspi et al. 2000).  As discussed by Krolik (2001), this simplistic approach
may be subject to a variety of unquantified systematic uncertainties.  
Nevertheless, Gebhardt et al. (2000b) and Ferrarese et al. (2001) have 
demonstrated, for a subset of the reverberation-mapped AGNs with available 
stellar velocity dispersions, that the virial BH masses, remarkably if 
unexpectedly, do roughly conform to the \msigma\ relation of inactive galaxies.
Interestingly, both NGC 4395 and POX 52 follow the low-mass
extrapolation of the Tremaine et al. (2002) fit to 
the \msigma\ relation (Filippenko \& Ho 2003; Barth et al. 2004; Ho 2004b).
The virial masses scatter around the \msigma\ relation with a spread of 
$\sim 0.5$ dex, which for the moment can be taken as an estimate of the 
systematic uncertainty of this method.

Certain aspects of our analysis may introduce additional errors into
the mass measurements, although we believe that they are small
relative, or at most comparable, to the intrinsic systematic
uncertainty discussed above.  There are two basic measurements that
affect our mass determinations, namely \lf\ and $\upsilon_{\rm FWHM}$.
Both of these quantities become increasingly challenging to measure
when galaxy contamination is large.  The complete profile of weak
emission lines with very broad wings becomes very difficult to measure
when the contrast between the AGN and the galaxy is low.  Dilution
from host galaxy starlight, which can never be removed perfectly, {\it
always}\ leads one to underestimate the true width of a weak,
low-contrast broad emission line.  This has been demonstrated most
conclusively for some objects by comparing starlight-subtracted
ground-based spectra with small-aperture spectra taken with the {\it
Hubble Space Telescope}\ (e.g., Ho et al. 2000; Shields et al.  2000):
the small-aperture spectra, largely free from starlight contamination,
reveal {\it much}\ broader \halpha\ wings than the ground-based
spectra.  To mitigate this problem, and that associated with measuring
\lf\ (\S 3.4), in this study we have chosen to concentrate on objects
where the AGN component is largely dominant.  This is evident from
Figures 1--4 by the overall weakness of stellar absorption features in
the spectra.

We note that substituting \halpha\ for \hbeta\ to measure
$\upsilon_{\rm FWHM}$ should introduce negligible additional
uncertainty into the mass estimates.  Although the physical conditions
in the BLR that give rise to \halpha\ and \hbeta\ emission in
principle may be slightly different (e.g., Kwan \& Krolik 1981), in
practice the two lines are observed to have rather similar velocity
profiles, with \hbeta\ on average being only slightly broader than
\halpha.  From a study of 19 Seyfert 1 galaxies, Osterbrock
\& Shuder (1982) find that the broad \hbeta\ and \halpha\ lines have
nearly identical full widths at zero intensity, and on average \hbeta\
is 16\% broader than \halpha\ at FWHM.  If this trend holds, then we
may be underestimating the mass in Equation 2 by $\sim$30\%, which is
insignificant compared to the much larger 0.5 dex uncertainty in the
virial mass zeropoint.  Moreover, in the two AGNs with
intermediate-mass BHs known to date---NGC 4395 and POX 52---there is
no indication that the line profile of broad \halpha\ differs
systematically from that of \hbeta\ (Filippenko \& Sargent 1989;
Kraemer et al. 1999; Barth et al.  2004).  We await future
measurements of stellar velocity dispersions to investigate potential
systematic differences in our virial mass estimates, but for now we
assume that the \halpha\ linewidth can be substituted for that of
\hbeta.

With the BH virial masses at hand, we are in a position to select the
intermediate-mass BH candidates.  With a cut-off of $10^6$~\msun, the
final sample contains 19 broad-line objects (Fig. 5).  Their
spectra are in Figures 1--4.  Table 1 shows the basic
properties of the sample, Table 2 lists measurements of
the emission-line parameters, and Table 3 gives the
luminosity and mass estimates.


\begin{figure}
\begin{center}
\epsfig{file=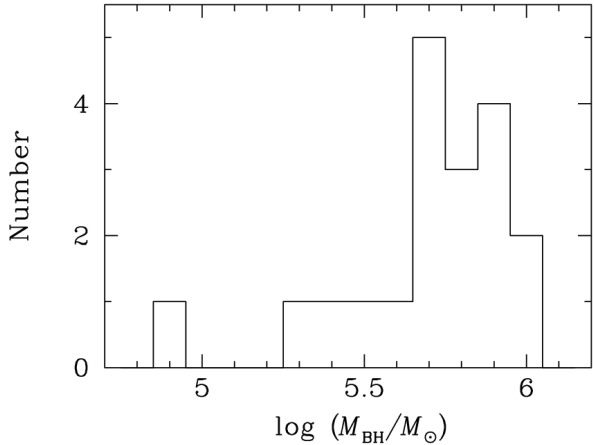,angle=-90,width=3.1in,keepaspectratio=true}

\vskip -0.1mm
\figcaption[f5.eps]{ 
The distribution of virial BH masses for our sample.
\label{fig5}}
\vskip -10mm
\end{center}
\end{figure}



\begin{deluxetable*}{rcccllcc}
\tablecolumns{8}
\tabletypesize{\scriptsize}
\tablewidth{0pc}
\tablecaption{Luminosity and Mass Measurements \label{tablemags}}
\tablehead{ 
\colhead{ID} & 
\colhead{$M_g$(total)} & 
\colhead{$M_g$(AGN)} & 
\colhead{$M_g$(host)} &
\colhead{$\beta$} &
\colhead{log $L_{5100}$} & 
\colhead{log $M_{\rm BH}$} &
\colhead{log $L_{\rm bol}/L_{\rm Edd}$} \\
\colhead{(1)} & 
\colhead{(2)} & 
\colhead{(3)} & 
\colhead{(4)} & 
\colhead{(5)} & 
\colhead{(6)} & 
\colhead{(7)} &
\colhead{(8)}
}
 
\startdata

1 &$-$19.1 & $-$17.8 & $-$18.7 &    0.60 & 43.06 & 5.86 & 0.092\\
2 &$-$18.9 & $-$15.4 & $-$18.9 &    1.00 & 41.87 & 4.92 & $-$0.15\\
3 &$-$18.7 & $-$17.4 & $-$18.4 &    0.11 & 42.91 & 5.84 & $-$0.03\\
4 &$-$18.6 & $-$17.1 & $-$18.3 &    0.88 & 42.80 & 5.72 & $-$0.02\\
5 &$-$19.2 & $-$17.7 & $-$18.9 &    0.27 & 43.10 & 5.74 & 0.254\\
6 &$-$18.8 & $-$17.7 & $-$18.3 &    0.076 & 42.93 & 5.86 & $-$0.03\\
7 &$-$18.8 & $-$17.1 & $-$18.5 &    0.44 & 42.97 & 5.91 & $-$0.04\\
8 &$-$20.4 & $-$18.3 & $-$20.2 &    0.14 & 43.35 & 5.77 & 0.475\\
9 &$-$20.8 & $-$18.4 & $-$20.6 &    1.00 & 43.02 & 5.69 & 0.223\\
10 &$-$19.7 & $-$18.4 & $-$19.4 &   0.46 & 43.33 & 5.93 & 0.296\\
11 &$-$19.5 & $-$17.5 & $-$19.3 &   0.76 & 43.22 & 5.75 & 0.362\\
12 &$-$19.9 & $-$18.1 & $-$19.7 &   0.79 & 43.17 & 5.98 & 0.082\\
13 &$-$20.6 & $-$18.1 & $-$20.5 &   0.39 & 43.32 & 5.99 & 0.218\\
14 &$-$19.7 & $-$15.4 & $-$19.7 &   1.00 & 41.92 & 5.30 & $-$0.48\\
15 &$-$18.8 & $-$18.3 & $-$17.8 &   0.24 & 42.97 & 5.71 & 0.153\\
16 &$-$19.1 & $-$17.4 & $-$18.9 &   1.00 & 42.53 & 5.58 & $-$0.15\\
17 &$-$18.9 & $-$17.6 & $-$18.5 &   0.31 & 43.05 & 5.35 & 0.592\\
18 &$-$20.3 & $-$18.0 & $-$20.1 &   1.00 & 43.09 & 5.74 & 0.240\\
19 &$-$19.0 & $-$15.0 & $-$19.0 &   1.00 & 41.69 & 5.54 & $-$0.95\\

\enddata	
\tablecomments{ 
Col. (1): Identification number assigned in this paper.
Col. (2): Total $g$-band absolute magnitude.
Col. (3): AGN $g$-band absolute magnitude, estimated from $L_{5100}$
given in 
Col. (6) and a conversion from $L_{5100}$ to $M_g$ assuming 
$f_{\lambda} \propto \lambda^{-\beta}$, where $\beta$ is given in
Col. (5);  typical uncertainty is $\sim 0.4$~mag.
Col. (4): Host galaxy $g$-band absolute magnitude, obtained by 
subtracting the AGN luminosity from the total luminosity.
Col. (5): Power-law slope $\beta$, where $f_{\lambda} \propto
\lambda^{-\beta}$; typical uncertainty is $\sim 10 \%$.  A value of 1.00 
indicates that it was fixed in the PCA decomposition.  
Col. (6): AGN continuum luminosity at 5100 \AA\ ($\mathrm{erg~s^{-1}}$), 
estimated from a power-law fit to the continuum; typical uncertainty is 
$\sim 10 \%$.  Col. (7): Virial mass estimate of the BH ($M_{\odot}$).
Col. (8): Ratio of the bolometric luminosity (see text) to the Eddington 
luminosity.
}
\end{deluxetable*}				   


\section{Discussion}

We employ a simple technique to exploit the breadth of the SDSS
galaxy spectroscopy to extend the demography of central BHs at least 1
order of magnitude below the $10^6$ \msun\ threshold, a regime hardly
explored previously.  We present the first sizable sample of AGNs with
candidate intermediate-mass BHs.  The sample contains 19 objects with
virial BH masses ranging from $M_{\rm BH} \approx 8\times 10^4$ to
$10^6$ \msun\ (Fig. 5).  We now briefly examine some of the
properties of these objects.

\subsection{Spectral Properties}

In the literature to date, the closest analog to our sample (aside
from NGC 4395 and POX 52) is the class of AGNs known as narrow-line
Seyfert 1s (NLS1), which are thought to possess relatively low-mass BHs emitting
at a high fraction of their Eddington rates (e.g., Boroson 2002).
Since our sample selection required a relatively prominent AGN
component relative to the host galaxy, the sample should
preferentially single out AGNs in a relatively ``high'' state.

First classified as a group by Osterbrock \& Pogge (1985), NLS1s are
typically characterized by broad permitted lines with FWHM $\lesssim
2000$ \kms, prominent \ion{Fe}{2} emission, and relatively weak
[\ion{O}{3}] $\lambda \lambda$4959, 5007 lines.  NLS1s also tend to
emit copiously in the soft X-ray band (e.g., Boller, Brandt, \& Fink
1996; but see Williams, Pogge, \& Mathur 2002), 
presumably as a consequence of their
high accretion rates (Pounds, Done, \& Osborne 1995).  Our sample
conforms to some, but not all, of these expectations.  By the
linewidth criterion, certainly all of the objects would technically
qualify as ``narrow-line'' sources: the broad \halpha\ component has
FWHM ranging from 464 to 1730 \kms, with an average value of 836 \kms.
However, we note that our sample shows a greater diversity of
\ion{Fe}{2} and [\ion{O}{3}] strengths than in previous samples of
NLS1s.  \ion{Fe}{2} emission is common, but not
ubiquitous\footnote{The same comment applies to NGC 4395 and POX 52,
both of which may be formally regarded as NLS1s.  Fe~{\tiny II}
multiplets can be seen in the UV spectrum of NGC 4395 (Filippenko et
al. 1993), but they are quite weak in its optical spectrum (Filippenko
\& Sargent 1989).  Fe~{\tiny II} emission is also not prominent in the
optical spectrum of POX 52 (Barth et al. 2004).}.  For example, the
intensity ratio of \ion{Fe}{2} $\lambda$4570 relative to total (broad
+ narrow) \hbeta, where \ion{Fe}{2} $\lambda$4570 is the complex
between 4434 and 4680 \AA, varies from $\sim$0.465 to 3.17, with a 
mean of $1.32 \pm 0.16$, calculated using the Kaplan-Meier product-limit
estimator to account for upper limits (Feigelson \& Nelson 1985).
This is to be
compared with $\langle$\ion{Fe}{2} $\lambda$4570/\hbeta(total)
$\rangle$ = $0.67\pm0.04$ for the 56 NLS1s presented 
by V\'eron-Cetty, V\'eron, \& Gon{\c c}alves (2001).  
Similarly, in
our sample [\ion{O}{3}]/\hbeta(total) = 0.335 to 12.0, with a mean of
$2.24 \pm 0.72$.  In contrast V\'eron-Cetty et al. (2001) find
$\langle$[\ion{O}{3}]/\hbeta(total)$\rangle$ = $0.65 \pm 0.09$, 
and Constantin \& Shields 2003 find
$\langle$[\ion{O}{3}]/\hbeta(total)$\rangle$ = 0.39 for their composite
spectrum of 22 NLS1s.
Thus, compared to classical NLS1s, we find stronger
[\ion{O}{3}] and \ion{Fe}{2} emission.
The distribution of power-law slopes for the AGN-dominated spectra are also 
quite interesting (Table 3).
While $1.5 < \beta < 1$ is a standard range for luminous AGNs (e.g.,
Malkan 1988), we
find an average value of $\beta = 0.42$.


\begin{figure*}
\begin{center}
\epsfig{file=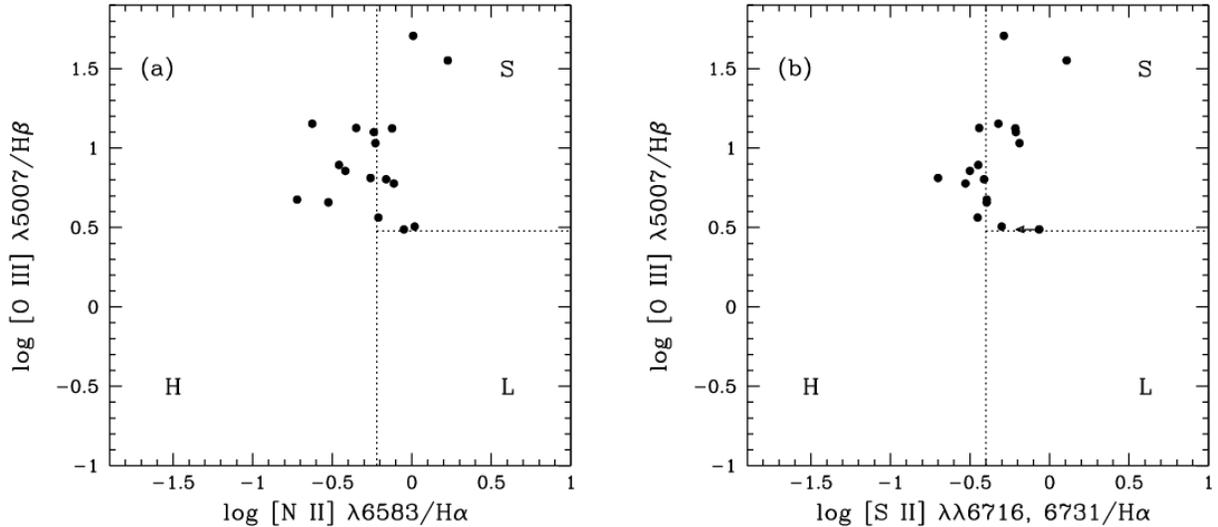,angle=-90,width=0.9\textwidth,keepaspectratio=true}
\vskip -0.1mm
\figcaption[f6.eps]{ 
Diagnostic diagram plotting log [O~{\tiny III}] $\lambda$5007/\hbeta\
versus log [N~{\tiny II}] $\lambda$6583/\halpha\ ({\it a}) and versus
log [S~{\tiny II}] $\lambda\lambda$6716, 6731/\halpha\ ({\it b}).  The 
line ratios have not been corrected for reddening, but this should not 
matter because of the close wavelength separation of the lines.  
Object \#15 was omitted because [N II], [S II], and narrow H$\alpha$ 
were not detected.  The dotted lines mark 
the boundaries of the three main classes of emission-line nuclei, according
to the convention of Ho et al. (1997a): ``H'' = H~{\tiny II} nuclei, ``S'' = 
Seyferts, and ``L'' = LINERs. 
\label{fig6}}
\vskip -5mm
\end{center}
\end{figure*}



\begin{figure}
\begin{center}
\epsfig{file=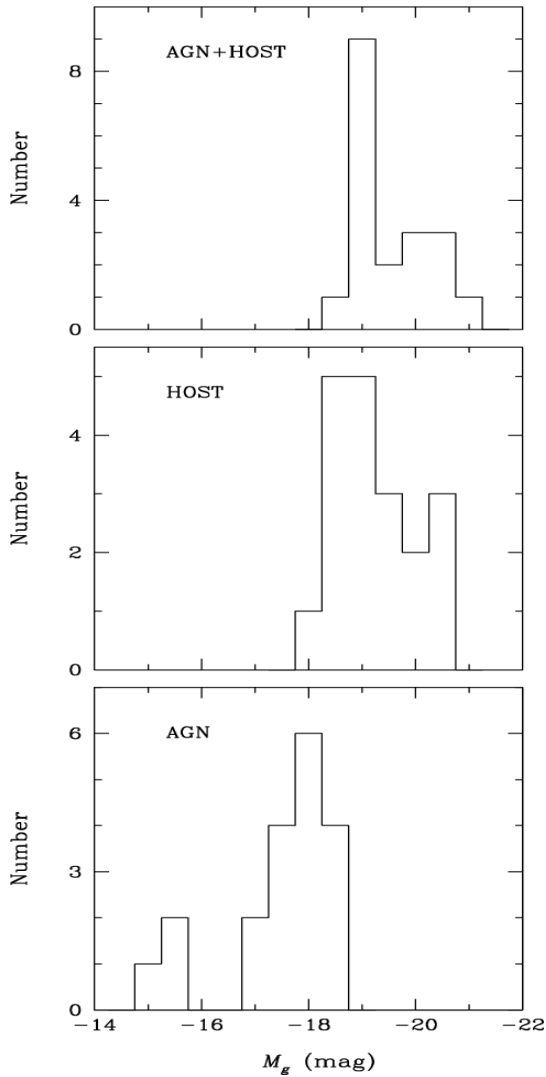,width=2.8in,height=5.6in} 
\vskip -0.1mm
\figcaption[f7.eps]{
Distributions of $g$-band absolute
magnitudes.  The magnitudes pertain to the entire system ({\it top}),
the host galaxy alone ({\it middle}), and the AGN alone ({\it bottom}).
The AGN contribution was estimated from \lf, which was then converted
to a $g$ magnitude assuming an $f_\lambda \propto \lambda^{-\beta}$
spectrum, where $\beta$ is shown in Table 3.
\label{fig7}}
\vskip -5mm
\end{center}
\end{figure}


\begin{deluxetable}{rcccc}
\tablecolumns{5} 
\tablewidth{0pc}
\tabletypesize{\scriptsize}
\tablecaption{ X-ray Detections\label{tablexr}}
\tablehead{ 
\colhead{ID} & 
\colhead{$C$} & 
\colhead{log $N_{\mathrm{H}}$} & 
\colhead{log $f_{\rm X}$} &
\colhead{log $L_{\rm X}$}\\
\colhead{(1)} & \colhead{(2)} & \colhead{(3)} &
\colhead{(4)} & \colhead{(5)} 
}

\startdata

1 & 0.0483 & 20.59 & $-$12.21 & 42.92\\
7 & 0.0276 & 20.59 & $-$12.46 & 42.86\\
8 & 0.113 & 20.37 & $-$12.86 & 42.32\\
12 & 0.100 & 20.30 & $-$12.92 & 42.51\\
14 & 0.0564 & 20.40 & $-$12.16 & 42.07\\
17 & 0.0174 & 20.50 & $-$12.66 & 42.70\\

\enddata	

\tablecomments{ 
Col. (1): Identification number assigned in this paper.
Col. (2): {\it ROSAT} count rate (counts s$^{-1}$).
Col. (3): Galactic column $N_{\rm H}$ (cm$^{-2}$), calculated
following Dickey \& Lockman (1990).
Col. (4): X-ray flux in the 0.5--2 keV band (\flux) assuming 
a power-law spectrum with $\Gamma = 3$ and $N_{\rm H}$ from Col. (3).
Col. (5): X-ray luminosity in the 0.5--2 keV band (erg~s$^{-1}$).
}

\end{deluxetable}


Only six of the 19 sources (30\%) were
detected in the {\it ROSAT} All-Sky Survey (Table 4),
confirming the finding of Williams et~al. (2002) that not
all NLS1s have detectable soft X-ray (0.5--2 keV) emission.  The data
are not of sufficiently high quality to make spectral analysis
possible, so to convert count rate to flux we assume a power-law
spectrum with a fixed photon index of $\Gamma$ = 3, which is roughly
the average value observed in previous {\it ROSAT}\ studies of NLS1s
(e.g., Boller et al. 1996; V\'eron-Cetty et al. 2001; Williams et
al. 2002).  We use the online program {\tt webPIMMS}\footnote{{\tt
http://heasarc.gsfc.nasa.gov/Tools/w3pimms.html}} to do the
conversion, assuming that the only source of absorption is due to the
Galaxy along the line of sight.  The derived soft X-ray luminosities
range from $\sim 10^{42.1}$ to $10^{42.9}$ \lum\ (Table
4), which probably can be attributed mostly to the AGN.
The host galaxies of our sample are relatively faint (\S 4.3), with 
absolute $g$-band magnitudes that roughly translate to $L_{B} \approx 
10^{9.4}-10^{10.1} \, L_{\sun}$ (Fukugita et al. 1996).  If these galaxies follow the
$L_{B}-L_{X}$ relation of normal galaxies (e.g., Fabbiano 1989), they are
expected to have $L_X \approx 10^{38.5}$ erg~s$^{-1}$.

Only one of the objects was detected in the VLA FIRST survey 
(Becker, White, \& Helfand 1995), although the entire sample
overlapped with the FIRST survey area.  Object \#10 has a 20~cm 
flux density of 1.3 mJy, which corresponds to a 
spectral power of $2.0 \times 10^{22}$ W Hz$^{-1}$.  
This level of radio emission is not atypical of normal galaxies (Condon 
1992), so we cannot conclude, without additional evidence, that it 
is associated with AGN activity.

Figure 6 places the sources on two commonly used
line intensity ratio diagnostic diagrams.
The dotted lines schematically denote the boundaries used by
Ho et al. (1997a) to demarcate the main classes of emission-line objects.
Although all the objects are clearly AGNs by the presence of their
prominent broad lines, some of them in the [\ion{O}{3}]/\hbeta\ vs.  \
[\ion{N}{2}]/\halpha\ and [\ion{O}{3}]/\hbeta\ vs.
[\ion{S}{2}]/\halpha\ diagrams formally lie outside of the AGN zone
([\ion{N}{2}]/\halpha\ $\geq$ 0.6 and [\ion{S}{2}]/\halpha\ $\geq$ 0.4).
This is rarely seen in local type 1 AGNs (Ho et al. 1997a).  Since the 
SDSS 
fibers subtend a significant fraction of each galaxy, the narrow 
emission-line ratios of the nuclei are, at least in part, contaminated by 
host galaxy emission, which would have the effect of lowering the
[\ion{S}{2}]/\halpha\ and [\ion{N}{2}]/\halpha\ ratios.  Thus, one
expects, a priori, that AGNs not selected by their narrow lines will,
in some cases, exhibit narrow-line ratios that fall below the AGN
threshold.

\subsection{Nuclear Luminosities and Eddington Ratios}

Since our sample was selected to have low BH mass, we expect that the
AGNs will generally be of relatively low luminosity.  Figure
7 shows the distribution of nuclear $g$-band absolute
magnitudes for the sample.  To estimate this quantity, we first
compute \lf\ (as described above),
which is then transformed to $M_g$.  For those spectra with direct
power-law fits, we use the best-fit spectral index, while for the
remainder we assume an $f_\lambda \propto \lambda^{-1}$
spectrum.  The AGN component has absolute magnitudes ranging from $M_g
\approx -15$ to $-18$ mag. Note that
these nuclear luminosities, though generally not as low as many local
Seyferts (Ho \& Peng 2001), are still significantly lower than in many
previously studied low-redshift type 1 AGNs (e.g., K\"{o}hler et al. 1997).

It is of interest to compare the nuclear luminosities with respect to
the Eddington luminosities.  The selection criteria---low BH mass
and a prominent AGN spectrum---are expected to favor objects with high
Eddington ratios.  A major uncertainty in this exercise is that we do
not have a robust estimate of the bolometric luminosity, but only a
measurement of the optical luminosity (\lf).  Since the spectral
energy distributions of AGNs are known to vary significantly with
accretion rate (Pounds et al. 1995; Ho 1999), it is doubtful that a
single bolometric correction is always applicable. Nonetheless,
assuming $L_{\rm bol}$ = 9.8\lf\ (McLure \& Dunlop 2004), the derived
Eddington ratios mostly cluster around $L_{\rm bol}/L_{\rm Edd}
\approx 1$ (Fig. 8; Table 3), as expected.


\begin{figure}
\begin{center}
\epsfig{file=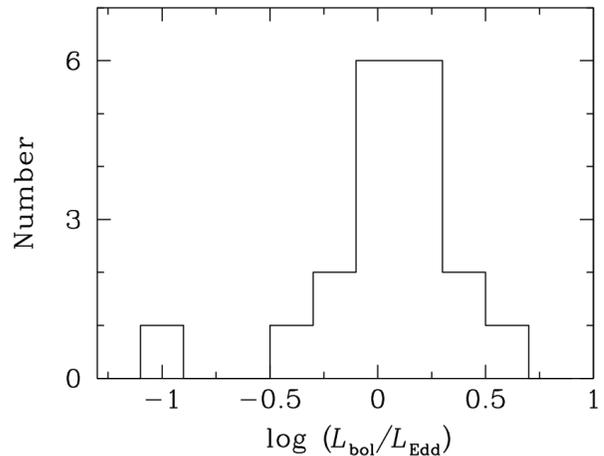,width=2.4in,keepaspectratio=true,angle=-90}
\vskip -0.1mm
\figcaption[f8.eps]{
Distribution of Eddington ratios.  The 
bolometric luminosity was estimated from \lf\ (see text).
\label{fig8}}
\vskip -5mm
\end{center}
\end{figure}


\subsection{Properties of the Host Galaxies}

The morphology of the host galaxies of intermediate-mass BHs is of
fundamental importance for understanding the origin of this class of
objects and their relationship to the overall demography of central
BHs in galaxies.  The two currently known nuclear intermediate-mass
BHs are found in anomalous hosts compared to the local AGN population
(Ho et al. 1997b, 2003; Kauffmann et al. 2003), which is invariably
affiliated with massive, bulge-containing galaxies.  Similarly,
essentially every galaxy known to contain a supermassive BH has a
bulge.  Yet, NGC 4395, which without a doubt lacks a bulge, evidently
hosts a nuclear BH (Filippenko \& Ho 2003).  The same could be said of
POX 52 (Ho 2004b).  Although POX 52 appears to be a dwarf elliptical
galaxy (Barth et al.  2004), this class of spheroids should be
regarded as physically distinct from classical bulges (Bender,
Burstein, \& Faber 1992), and they may have originated as tidally
stripped late-type disk galaxies (e.g., Moore et al. 1996).  Thus,
both NGC 4395 and POX 52 provide compelling evidence that a classical
bulge is {\it not}\ necessary for the formation of a central BH.

We have attempted to estimate the luminosities of the host galaxies by
subtracting the contribution of the AGN from the total (Petrosian)
$g$-band magnitude of the AGN plus host.  As illustrated in Figure 7 
(see also Table 3), in most cases the
luminosities of the AGNs are sufficiently low that they hardly perturb
the distribution of total magnitudes; thus, the host galaxy magnitudes
should be quite robust. Interestingly, the hosts are relatively
low-luminosity and therefore more likely to be
late-type galaxies: $M_g$(host) ranges
from $-17.8$ to $-20.6$ mag, with an average value of $-19.2$.  According to
Blanton et al.  (2003), the luminosity function of galaxies at $z =
0.1$ has $M_g^* = -19.39 + 5 \log h$.  For $h = 0.72$, as assumed
here, $M_g^* = -20.1$ mag.  Thus, the hosts of our sample are $\sim
1$ mag fainter than $M_g^*$.

Our selection technique is unbiased with respect to galaxy morphology.
Unfortunately, we have very limited information on the morphological
types of the host galaxies.  The low resolution of the SDSS imaging
makes morphological classification difficult for these objects.  By
selection, the galaxy centers are dominated by an AGN core, which makes
it impossible to say anything meaningful about the central structure
of the galaxies.  If the low BH masses are accompanied by a low-mass
(low-luminosity) galaxy or bulge, this makes it doubly hard to detect.
We show a selection of the more interesting images in Figure
9.  Based on the SDSS imaging, one-third of the galaxies
clearly have a disklike component.  Of the six that show disks, five
have significant galaxy light in their spectra, which required PCA
decomposition.  Given the low luminosities of these galaxies, perhaps
they are analogs of NGC 4395.  The remainder are barely resolved
enough to tell, but could be consistent with compact spheroids.
We caution that the images are of low
resolution and sensitivity, so in many cases we may be insensitive to
an extended stellar component.  Higher resolution, deeper imaging, for example
with the {\it Hubble Space Telescope}, is essential to determine the
true morphology and detailed structural parameters of these hosts.

In lieu of resolved morphological information, the SDSS collaboration
has developed statistical proxies that describe morphological trends
(Shimasaku et al. 2001; Strateva et al. 2001).  We consider the
``eclass'' (a PCA statistic), the \sur\ color, and the ``inverse
concentration index.''  The eclass ranges from $-0.35$ for early-type
galaxies to +0.5 for late-type galaxies. The mean eclass of our
sample is $0.24$.
This statistic is difficult to interpret because of the strong
contamination by the AGN.  For $\sim 150,000$ SDSS galaxies, Strateva et
al. (2001) find a bimodal distribution in \sur\ color, with the
division at \sur\ = 2.22 mag, which roughly divides early and late
types.  Our galaxies have a mean color of $u-r \approx 1.51$,
consistent with early-type galaxies.  An inverse concentration index
of $\sim 0.3$ corresponds to a de~Vaucouleurs (1948) $r^{1/4}$
profile, while an exponential profile has an index of $\sim 0.43$.
The mean inverse concentration index of our objects is 0.41.


\begin{figure}
\begin{center}
\epsfig{file=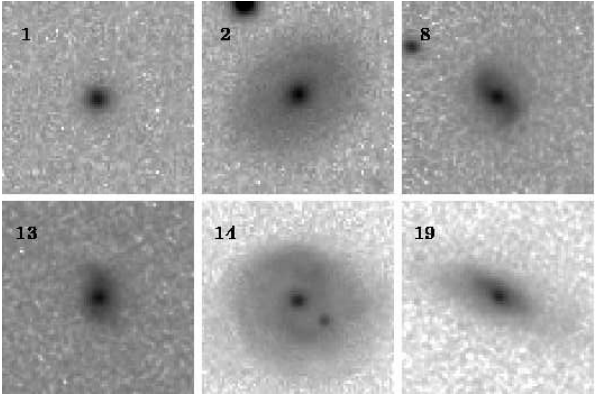,angle=-90,width=3.1in,keepaspectratio=true}
\vskip -0.1mm
\figcaption[f9.eps]{
Sample SDSS $g$-band images of six objects.  
The vast majority of the sample are barely resolved and 
look similar to object \#1.  In all images North is up and East is
to the left.
Each panel is $30^{\prime\prime} \times 30^{\prime\prime}$. 
\label{fig9}}
\vskip -5mm
\end{center}
\end{figure}


\subsection{Implications and Future Work}

While the discovery of the \msigma\ relation established a compelling
link between the evolution of galaxies and central BHs, models of this
process are faced with significant uncertainties.  Possible models
include seed BHs formed from Population III stars, which then gain
mass through mergers (Schneider et al. 2002; Volonteri, Haardt, \&
Madau 2003) or through accretion (Adams, Graff, \& Richstone 2001;
Islam, Taylor, \& Silk 2003).  Alternatively, seed BHs may form
directly from a collapsing gas cloud (Koushiappas, Bullock, \& Dekel
2004; for a more complete discussion, see the review by van~der~Marel
2004).  However, there are few direct observational constraints on the
nature of seed BHs, the path by which nuclear BHs grow, or the form of
the BH mass function.  The low end of the BH mass function, in
particular, provides critical input to check theoretical models of
quasar formation (e.g., Haehnelt, Natarajan, \& Rees 1998; Bromley,
Somerville, \& Fabian 2004), and it affects predictions of future
gravitational wave experiments like {\it LISA}\ (Hughes 2002).

A useful approach, first proposed by So{\l}tan (1982), matches the 
local BH mass density inferred from velocity dispersion measurements with
an accretion history inferred from observations of luminous quasars. Yu \& 
Tremaine (2002) conclude that, for standard assumptions about radiative 
efficiency and quasar lifetimes, accretion during the luminous quasar phase
can account for the majority of the local BH mass density.  Aside from the 
possibility of significant obscured accretion (Fabian 2004), this method is 
limited by its use of the SDSS velocity dispersion function, which at 
the moment is only well determined for early-type galaxies with 
$\sigma_{\star} \gtrsim 200$ \kms\ (Sheth et al. 2003), corresponding 
to $M_{\rm BH} \gtrsim 10^8$ \msun.  

Much of the current observational and theoretical effort on
intermediate-mass BHs has focused on studies of star clusters
(Gebhardt et al. 2002; Gerssen et al. 2002) and so-called
ultraluminous X-ray sources (see van~der~Marel 2004).  The existence
of BHs in either class of objects is still far from proven, and, in
any event, their relationship to galactic nuclei is unclear.  With the
recognition that galaxies like NGC 4395 and POX 52 do harbor central
intermediate-mass BHs, the most straightforward way to make progress
would be to find many more objects like them.  This study represents
the first step in this direction.  By sifting through the massive SDSS
database, we have identified a sample of AGNs that we believe are
promising candidates for hosting BHs with masses below the $10^6$
\msun\ threshold.

Our sample opens up several new avenues for further study.  While
well-defined, our sample is likely to have missed intermediate-mass BH
candidates such as NGC 4395 or POX 52, whose spectrum is dominated by
narrow emission lines, with the broad-line component visible only
under very high signal-to-noise ratio.  Moreover, if either AGN were
embedded in a much more luminous host galaxy, the weak broad lines
would be even harder to discern.  Furthermore, we know that pure
narrow-line (type 2) AGNs do exist, either because the BLR is obscured
or intrinsically absent, and there is no reason to suppose that this
situation does not persist to the regime of intermediate-mass BHs.  A
complimentary approach, then, is to use the width of the narrow emission
lines to select potential candidates.  While the detailed geometry and
kinematics of the narrow-line region (NLR) in AGNs are complex and not
well understood, it has been argued that its velocity field is largely
dominated by the gravitational potential of the bulge of the host
galaxy (Smith, Heckman, \& Illingworth 1990; Whittle 1992a, b).  If
so, then the velocities in the NLR are roughly virial, and on average
the velocity dispersion of the line-emitting gas, $\sigma_{\rm g}$,
should be approximately equal to the velocity dispersion of the bulge
stars, \sigmastar.  The most detailed comparison between NLR and
stellar kinematics for a large sample of Seyfert galaxies has been
done by Nelson \& Whittle (1996), who indeed find that in most cases
$\sigma_{\rm g}\, \approx$ \sigmastar, where $\sigma_{\rm g}$ was
measured using [\ion{O}{3}] $\lambda5007$.  The exception are objects
with powerful, jetlike radio structures where $\sigma_{\rm g}$ can be
much larger than \sigmastar, presumably because of extra
nongravitational motions induced by the radio outflows.  Thus, in
general one expects that $\sigma_{\rm g}$, as measured in [\ion{O}{3}]
$\lambda5007$ or other optical forbidden lines, can be taken as a
rough proxy, or perhaps more realistically as an upper limit, for
\sigmastar.  We are in the process of enlarging the current sample
through a selection based on this strategy.

Another
important next step is to obtain stellar velocity dispersions for
these objects.  Such measurements will test whether the BH mass
estimates we derive obey the \msigma\
relation.  Conversely, if we assume that the \msigma\ relation is
universally valid, the stellar velocity dispersion measurements will
test the reliability of the virial mass estimator for AGNs. 
It would also be valuable to compare the velocity dispersions derived
from stars with those derived from gas.  Another area requiring
attention is to obtain high-resolution, deep imaging.  Much better data are
needed to quantify the morphologies and detailed structural parameters
of the host galaxies.  Do these objects have bulges, and if so, what
kind and where do they fall on galaxy scaling relations like the
fundamental plane?  Finally, much work remains to be done, especially
at non-optical wavelengths, to further characterize the properties of the AGNs
themselves.

\acknowledgements 
J.~E.~G. is funded by a National Science Foundation
Graduate Research fellowship.  L.~C.~H. acknowledges support by the
Carnegie Institution of Washington and by NASA grants from the Space
Telescope Science Institute (operated by AURA, Inc., under NASA
contract NAS5-26555).  We are grateful to Lei Hao for making available
her PCA software, to Aaron Barth and John Huchra for helpful
discussions on various aspects of this work, to Pat Hall for his
timely and helpful referee's report, and to the entire SDSS
collaboration for providing the extraordinary database and processing
tools that made this work possible.  This research has made use of
data obtained from the High Energy Astrophysics Science Archive
Research Center (HEASARC), provided by NASA's Goddard Space Flight
Center.

\end{document}